\documentclass[iop,tighten,apj]{emulateapj}

\slugcomment{Accepted for publication in ApJ}

\begin{document}

\title{Tiny scale opacity fluctuations from VLBA, MERLIN and VLA observations of H~{\sc i} absorption toward 3C~138}

\author{Nirupam Roy\altaffilmark{1,2}, Anthony H. Minter\altaffilmark{3}, W. M. Goss\altaffilmark{1}, Crystal L. Brogan\altaffilmark{4} and T. J. W. Lazio\altaffilmark{5}}

\altaffiltext{1}{National Radio Astronomy Observatory, 1003 Lopezville Road, Socorro, NM 87801, USA}
\altaffiltext{2}{Contact author's electronic address: nroy@aoc.nrao.edu (NR). NR is a Jansky Fellow of the National Radio Astronomy Observatory (NRAO). The NRAO is a facility of the National Science Foundation (NSF) operated under cooperative agreement by Associated Universities, Inc. (AUI).}
\altaffiltext{3}{National Radio Astronomy Observatory, P.O. Box 2, Green Bank, WV 24944, USA}
\altaffiltext{4}{National Radio Astronomy Observatory, 520 Edgemont Road, Charlottesville, VA 22903, USA}
\altaffiltext{5}{Jet Propulsion Laboratory, California Institute of Technology, Pasadena, CA 91109, USA}

\begin{abstract}
The structure function of opacity fluctuations is a useful statistical tool to 
study tiny scale structures of neutral hydrogen. Here we present high 
resolution observation of H~{\sc i} absorption towards 3C~138, and estimate 
the structure function of opacity fluctuations from the combined VLA, MERLIN 
and VLBA data. The angular scales probed in this work are $\sim 10-200$ 
milliarcsec (about $5-100$ AU). The structure function in this range is 
found to be well represented by a power law $S_{\tau}(x) \sim x^\beta$ with 
index $\beta \sim 0.33 \pm 0.07$ corresponding to a power spectrum $P_{\tau}(U) 
\sim U^{-2.33}$. This is slightly shallower than the earlier reported power law 
index of $\sim 2.5-3.0$ at $\sim 1000$ AU to few pc scales. The amplitude of 
the derived structure function is a factor of $\sim 20-60$ times higher 
than the extrapolated amplitude from observation of Cas~A at larger scales. On 
the other hand, extrapolating the AU scale structure function for 3C~138 
predicts the observed structure function for Cas~A at the pc scale correctly. 
These results clearly establish that the atomic gas has significantly more 
structures in AU scales than expected from earlier pc scale observations. 
Some plausible reasons are identified and discussed here to explain these 
results. The observational evidence of a shallower slope and the presence of 
rich small scale structures may have implications for the current 
understanding of the interstellar turbulence.
\end{abstract}


\section{Introduction}
\label{sec:int}

The interstellar medium (ISM) is known to have clumpy density and
velocity structures. Scale-free intensity fluctuations in a variety of
tracers (e.g.  H~{\sc i} 21~cm emission, dust emission) is detectable
over a wide range of scales. This scale-free fluctuation is
generally interpreted as the signature of a turbulent ISM. The
evidence for small scale structures and turbulence in the atomic ISM
of the Galaxy have been found through various observational
techniques. Power law scaling for the velocity dispersion of the H~{\sc i} 
21~cm emission has been observed to sub-pc scales \citep{nr08}. On
somewhat larger scales, the intensity fluctuations in H~{\sc i} 21~cm
emission show a scale free power spectrum
\citep{cro83,gre93,dic01}. The H~{\sc i} opacity towards Cassiopeia A
and Cygnus A show fluctuations on scales ranging from $500$ AU to $4$
pc for the absorption in the Perseus arm, the local arm, and the Outer
arm \citep{ddg00,nr10}. Very long baseline interferometry (VLBI)
observations reveal structures on scales as small as $25$ AU
\citep{die76,dia89,dav96,fai98,fai01,bro05}. \citet{des92},
\citet{fra94}, \citet{joh03} and \citet{sta10} have shown H~{\sc i}
opacity variations on scales of $5-100$ AU from multi-epoch
observations of some high velocity pulsars. \citet{dha00} have
reported similar opacity variations on $40-1000$ AU scales towards the
microquasar GRS~$1915+105$. Observations of some other pulsars,
however, do not show any opacity variations
\citep{joh03,sta03,min05,sta10}, indicating that such tiny scale
structures may be rare.

Small scale ISM structures are not unique to the Milky Way. The H~{\sc
  i} emission from the Large Magellanic Cloud, the Small Magellanic
Cloud, several dwarf galaxies in the M81 group and a sample of dwarf
and nearby spiral galaxies also show considerable small scale
intensity fluctuations
\citep{sta99,wes99,sta01,elm01,beg06,kim07,dut09a,dut09b}.

From an observational point of view, the H~{\sc i} 21~cm opacity fluctuations 
depend on fluctuations in the density and temperature of the gas. But, for 
small density fluctuations, the dependence on temperature is weak 
\citep{ddg00}. Earlier, these small fluctuations were believed to be physical 
structures with densities $\sim 10^4-10^5$~cm$^{-3}$. However, explaining 
the existence of these structures in equilibrium with orders of magnitude 
lower density gas in the surrounding medium, is challenging. Later, it was 
suggested that the fluctuations at larger spatial scales could, in projection, 
give rise to the observed small scale transverse fluctuations \citep{des00}. 
On the other hand, there are some direct observational evidence of $\sim 3000$ 
AU size H~{\sc i} structures with a density of $\sim 100$~cm$^{-3}$ 
\citep{bra05,sta05}. Recent numerical simulations of the turbulent ISM have 
also indicated the existence of tiny scale structures 
\citep{nag06,vaz06,hen07}. The short evaporation time-scale ($\sim 1$ Myr) of 
these structures imply that they can survive if either the ambient pressure is 
much higher than the average ISM pressure or they are formed on a similar 
time-scale. In any case, their physical properties are not yet well understood.

Here we present results on tiny scale H~{\sc i} opacity fluctuations
towards the quasar 3C~138 derived from VLBA, MERLIN and VLA
data. Combining MERLIN and VLA data with VLBA data has resulted in
improved sensitivity and larger spatial dynamic range to detect small
optical depth fluctuations and to derive the structure function. The
data analysis method is outlined in Section \S\ref{sec:oaa}.  Section
\S\ref{sec:rsl} contains the results, and the conclusions are
presented in Section \S\ref{sec:con}.

\section{Data analysis}
\label{sec:oaa}

For this analysis, we have used VLBA, MERLIN and VLA data. Results
from the MERLIN and VLBA observations were previously presented by
\citet{dav96}, \citet{fai98} and \citet{bro05}. The VLBA observation
(Project BD0026) was carried out on September 10, 1995. The spectral
resolution of this data is $\sim 0.4$~km~s$^{-1}$. The MERLIN
observation was carried out on October 22 and 23, 1993, and the
spectral resolution in this case is $\sim 0.4$ km~s$^{-1}$. The VLA
A-configuration observations were carried out on September 07
and 13, 1991 (Project AS0410 and TEST). The channel width and spectral
resolution in this case are $\sim 0.8$ and $\sim 0.6$~km~s$^{-1}$,
respectively. \citet{bro05} reported temporal variation of H~{\sc i}
opacity towards 3C~138 from the VLBA observations taken in 1995, but
also include data from 1999 and 2002. Here, we have only included the
1995 data, since it is closest to the MERLIN and VLA observation
epochs. The longest baseline was about $0.18$, $1.0$ and $36~ {\rm
  M}\lambda$ (angular resolution of about $1000$, $200$ and $6$ mas)
for the VLA, MERLIN and VLBA data respectively. However, even with
the combination of the VLBA, MERLIN and VLA, sufficient brightness
temperature sensitivity was achieved only on angular scales larger
than about $20$~mas, corresponding to baselines shorter than about
$\sim 15~ {\rm M}\lambda$.

\begin{figure}
\begin{center}
\includegraphics[scale=0.33, angle=-90.0]{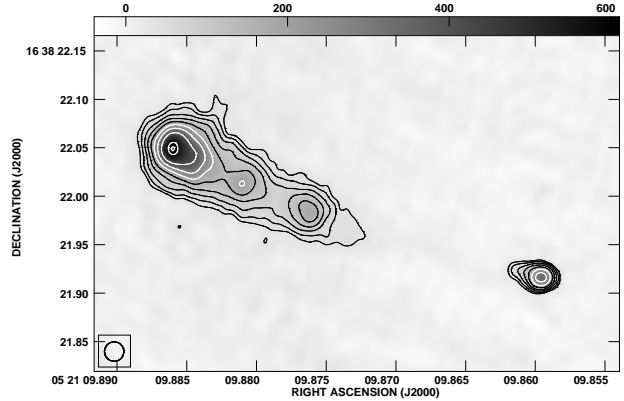}
\caption{\label{fig:ths1} $1.4$ GHz continuum image of 3C~138 from the
  combined VLBA, MERLIN and VLA data with a $20$ mas restoring
  beam. The rms noise of the image is $12$ mJy/beam. The contour
  levels at (1.0, 1.4, 2.0, 2.8, 4.0, 5.6, 8.0, 11.2, 16.0, 16.8)
  $\times 36$ mJy/beam are overplotted on the gray scale image with
  linear intensity scale.}
\end{center}
\end{figure}

Data analysis was carried out using the NRAO Astronomical Image
Processing System ({\small AIPS}). All the data were converted to
J2000.0 epoch, and the spectral axis was regrided to a common spectral
resolution of $0.4$~km~s$^{-1}$. A combined dataset was produced after
first reweighting the VLBA, MERLIN and VLA data to have similar data
weights (i.e. so they would contribute equally at the imaging stage).
A continuum and continuum-free dataset were produced by fitting the
line-free channels in the uv-plane. Self-calibration was repeatedly
done on the continuum data starting with a VLBA image as the initial
model, and then including the MERLIN and the VLA baselines at later
stages. The self-calibration solutions obtained for the continuum data
were also applied to the spectral line data cube. The final continuum
image, shown in Figure (\ref{fig:ths1}), and the spectral line image
cube were made using multiscale imaging with a $20$ mas restoring
beam. The structures of the continuum and line images agree with the
VLBA images of \citet{bro05}. The spectral image cube and the
continuum image were used to obtain the optical depth image cube. A
$3\sigma$ cutoff was used for blanking the continuum image, and the
same blanking window was applied to the spectral data to get rid of
possible contribution from high noise or other low level image
artifacts. The optical depth image cube was used to derive the opacity
fluctuation structure function.

We note that the structure function amplitude of the line-free channels is 
more than an order of magnitude lower than the structure function amplitude 
of the channels with H~{\sc i} absorption signal. But, for
the channels with H~{\sc i} signal, the optical depth noise is
significantly higher than that of the line-free channels. Since the
rms noise in the continuum image ($\sigma_C$) is significantly lower
than the rms noise in line channels ($\sigma_I$), the optical depth
noise can be approximated as $\sigma_\tau \approx e^\tau
(\sigma_I/I_0)$, where $I_0$ is the continuum image flux
density. Thus, for high optical depth, the noise is scaled up by a
factor of $e^\tau$. So, the optical depth noise is highly correlated with the
optical depth itself. Numerical analysis
shows that, as a result, the noise bias will have a similar power law
dependence on all angular scales, and will scale up the structure
function amplitude without changing the power law index. Moreover, the 
H~{\sc i} brightness temperature for the line channels will contribute to the 
system temperature, and will also increase the image noise. As a result, 
estimating the noise contribution from the line-free channels will 
underestimate the errors to the structure function due to optical depth 
uncertainties. 

Statistical errors on the derived structure function for individual
channels are estimated from the variance of the measurements within
the bin, and do not account for the high optical depth measurement
uncertainty. If the high opacity pixels are excluded, the effect of
this uncertainty and the noise bias to the derived structure function
is minimized. But in that case, the opacity image is modified by a
window function, and the structure function can only be derived for
about a factor of $4$ smaller range of angular scales, and only $\sim
8-10$ spectral channels. Hence, first the high opacity pixels are
included to derive the initial structure function. With significant
noise bias, the amplitude of the structure function in this case is
only an upper limit.  However, one can use a larger range of angular
scale to fit the power law index. Then $\tau/\sigma_\tau$ cutoff is
used to minimize the noise bias and to compute the amplitude of the
structure function from a limited range of angular scales. For both
cases, statistical errors are estimated using the channel to
channel variation of the structure function. Essentially, the opacity
values in any channel may be considered as an independent realization
drawn from a parent distribution of opacities with associated
uncertainty.  So, the channel to channel variation of the structure
function will include the opacity uncertainty effect. The errors are
estimated assuming that there is no spectral variation of the
structure function. Thus, one can only derive an average structure
function from all the spectral channels.

\section{Results}
\label{sec:rsl}

\begin{figure}
\begin{center}
\includegraphics[scale=0.33, angle=-90.0]{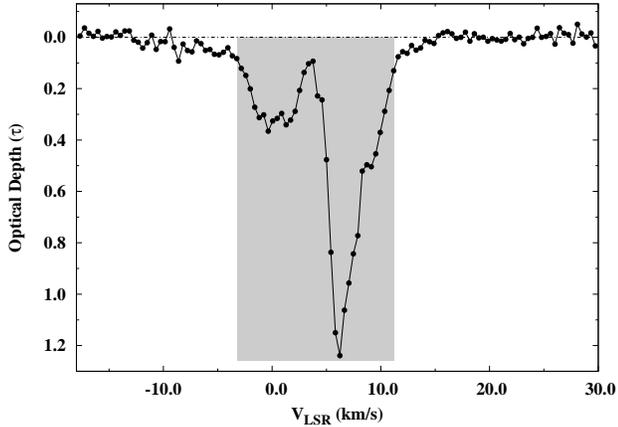}
\caption{\label{fig:ths2} Integrated H~{\sc i} 21~cm absorption
  spectrum towards 3C~138 from combined VLBA, MERLIN and VLA data. The
  channels in the shaded region are used to derive the structure
  function.}
\end{center}
\end{figure}

3C~138 is unresolved at the short baselines ($\leq 45 {\rm k}\lambda$)
of the VLA. Thus, by combining the VLBA, MERLIN and VLA data, we are
able to image 3C~138 maintaining the small scale structures at the
VLBA resolution, and, at the same time, restoring the total flux
density observed by the VLA. The intermediate MERLIN and VLBA
baselines provide a wide range of uv-coverage, allowing us to probe
structures at scales ranging from about ten milliarcsec to about
$0.20$ arcsec. Adopting a distance to the absorbing gas of 500 pc
\citep{fai98}, the range of angular scales translates to $5-100$
AU. The integrated H~{\sc i} 21~cm absorption spectrum is shown in
Figure~(\ref{fig:ths2}). The shaded region in Figure~(\ref{fig:ths2})
indicates the channels that have adequate signal to noise in optical
depth images for further analysis of small scale opacity
fluctuations. We have calculated the structure function for $36$
individual spectral channels, each with a velocity width of $\sim 0.4$
km~s$^{-1}$, and used all the measurements together to derive the best 
estimate of the structure function slope and amplitude.

For quantitative study of small scale fluctuations, statistical tools
like the correlation function, structure function, and power spectrum
are very useful.  The observed shape of these functions are expected
to be related to the physical processes that gives rise to the
fluctuations. The power spectrum of the opacity fluctuations
$P_{\tau}(U)$ is the Fourier transform of the correlation function
$\xi_{\tau}(x)$:
\begin{equation}
P_{\tau}(U) \equiv P_{\tau}(u,v) = \int\int\xi_{\tau}(l,m)e^{-2\pi i(ul+vm)}dldm \,,
\end{equation}
where $(l,m)$ and $(u,v)$ are the sky coordinate and inverse angular separation 
respectively. Here, $\xi_{\tau}(x)$ is defined as
\begin{equation}
\xi_{\tau}(x) \equiv \xi_{\tau}(l-l^{\prime},m-m^{\prime}) = \langle \delta\tau(l,m)\delta\tau(l^{\prime},m^{\prime})\rangle \ \,,
\end{equation}
and the structure function $S_{\tau}(x)= [\Delta\tau(x)]^2$ is given by 
\begin{equation}
S_{\tau}(x) \equiv S_{\tau}(l-l^{\prime},m-m^{\prime}) = \langle [\tau(l,m)-\tau(l^{\prime},m^{\prime})]^2\rangle \,.
\end{equation}
Strictly speaking, $P_{\tau}$ and $\xi_{\tau}$ are function of $\vec{U}$ and 
$\vec{x}$, respectively, but we have assumed statistical isotropy for the 
purpose of simplification. Thus, throughout this analysis, $P_{\tau}$ and 
$\xi_{\tau}$ are considered to only be functions of scalar amplitude $U$ and 
$x$, respectively. Generally, if the power spectrum is a power law of the form 
$P_{\tau}(U) = P_0U^{-\alpha}$ with $2 < \alpha < 4$, the structure function 
$S_{\tau}(x)$ will have a power law index of $\alpha-2$, and the rms opacity 
fluctuation $\Delta\tau(x)$ will have a power law index of $(\alpha-2)/2$ 
\citep{lnj75,des00}. We also note that if the density fluctuations are 
small (that is $\delta\rho/\langle\rho\rangle << 1$), the density fluctuation 
power spectrum is expected to be very similar to the opacity fluctuation power 
spectrum \citep{ddg00}.

\begin{figure}
\begin{center}
\includegraphics[scale=0.33, angle=-90.0]{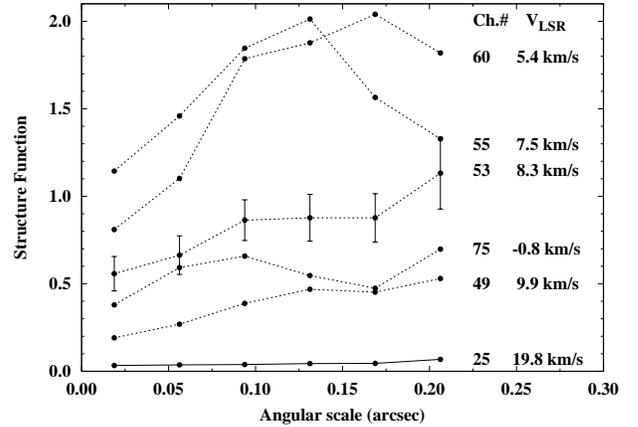}
\caption{\label{fig:ths3} Structure function from a few selected
  channels with H~{\sc i} absorption (dashed line), and from one
  channel with no absorption (solid line). For clarity, error bars are
  shown for only one channel.}
\end{center}
\end{figure}

\begin{figure*}
\begin{center}
\includegraphics[scale=0.67, angle=-90.0]{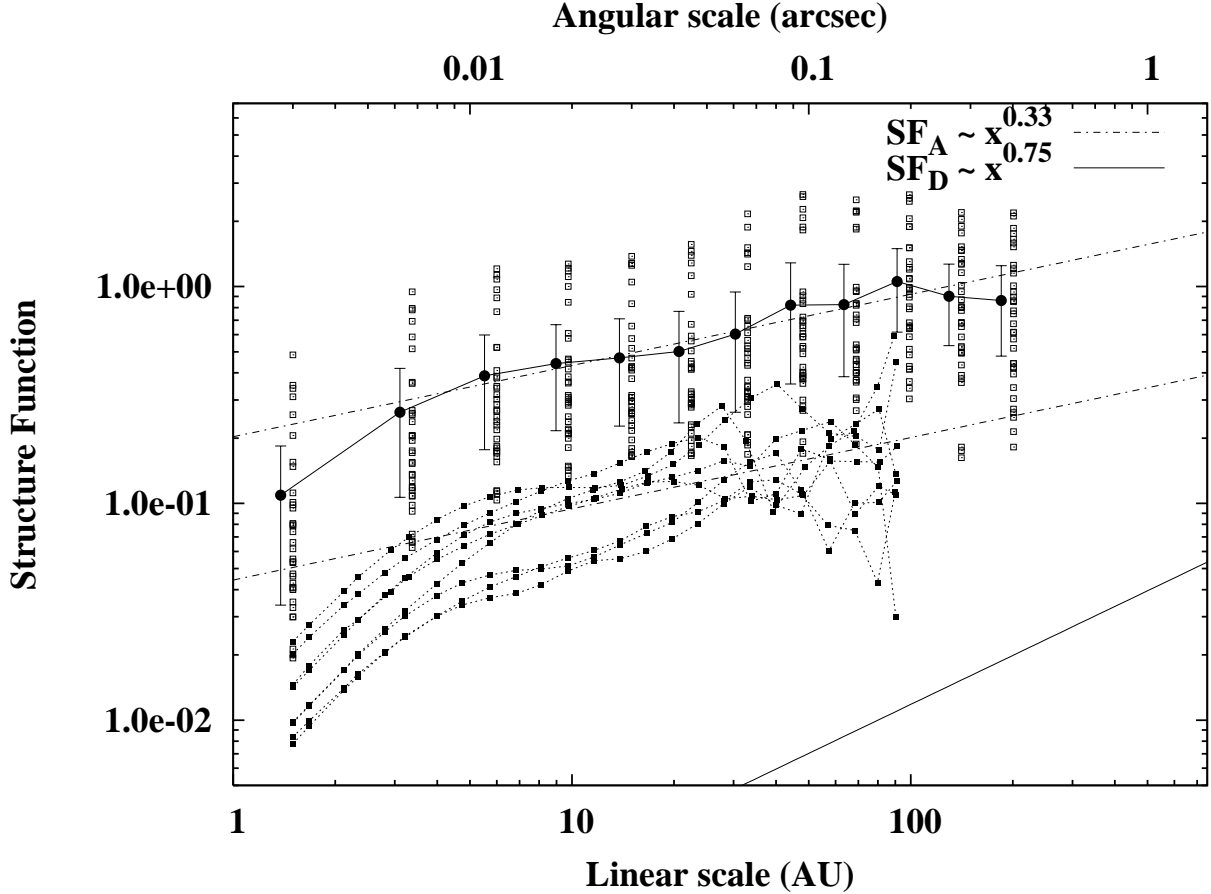}
\caption{\label{fig:ths6} Measured opacity fluctuation structure
  function from all spectral channels with adequate signal to
  noise. Open squares are estimates without a $\tau/\sigma_\tau$
  cutoff. The mean and rms values in each bin are shown as filled 
  circles with error bars (with position shifted slightly to
  the left for clarity). The best fit power law (shown as dash-dot lines) 
  for the ensemble has a power law index of $0.33 \pm 0.07$ ($3\sigma$). 
  The structure function derived using $\tau/\sigma_\tau = 2$ cutoff 
  (filled squares joined by dotted lines) is also consistent with a power law index of $0.33$. The solid line is the
  extrapolated structure function from \citet{des00}.}
\end{center}
\end{figure*}

The derived structure function of the H~{\sc i} absorption towards
3C~138 for a few spectral channels are shown in
Figure~(\ref{fig:ths3}). The amplitude and the shape of the function
changes from channel to channel. The structure function from one
spectral channel with no H~{\sc i} signal is also shown in this figure
for a comparison. The measured structure function for all 36 spectral
channels with adequate signal to noise is shown in Figure
(\ref{fig:ths6}). The open point symbols are structure function values
without any $\tau/\sigma_\tau$ cutoff. Clearly, below $\sim 10$
milliarcsec, the structures are smoothed by the synthesized
beam. Above $\sim 200$ milliarcsec, the effects of the spatial window
due to the shape of the continuum emission significantly modify and
flatten the measured structure function. The mean and rms values of
these points in each bin are shown as filled circles with error bars
(with position shifted slightly to the left for clarity). The best fit
power law for the ensemble (dash-dot lines in the figure) has a power
law index of $0.33 \pm 0.07$ ($3\sigma$), which corresponds to a power
spectrum with a power law index $\alpha = 2.33$.

Next, we have used a cutoff based on $\tau/\sigma_\tau$ to minimize
the noise bias and obtain a better estimate of the structure function
amplitude. As mentioned earlier, with this cutoff we have a reliable
estimate of the structure function only for $\sim 8-10$ channels and a
limited range of angular scales. Here, we have used a cutoff at
$\tau/\sigma_\tau = 2$ with usable range of $\sim 10-40$ mas for eight
spectral channels. These estimates (shown as filled points joined
by dotted lines in the same figure) are also consistent with a power law 
index of $0.33$. Since $\tau/\sigma_\tau \geq 2$,
the signal structure function is at least a factor of $4$ higher than
the noise bias case. So, the amplitude of the signal structure
function is constrained to be higher than $0.8$ times the observed
amplitude.  Unfortunately, for any higher cutoff, the flattening due to
the spatial window dominates, and essentially makes it impossible to
extract the true structure function. The solid line in the figure is
the structure function from \citet{des00} extrapolated from the
observed structure function power index of $0.75$ in the range of
$\sim 0.02-4.0$ pc. Clearly the observed structure function amplitude
at smaller scales ($5-100$ AU) is $\sim 20-60$ times higher than the
extrapolated value. In terms of rms opacity fluctuation, the observed
values are a factor of $4-8$ times higher than the predicted values.

\begin{figure}
\begin{center}
\includegraphics[scale=0.33, angle=-90.0]{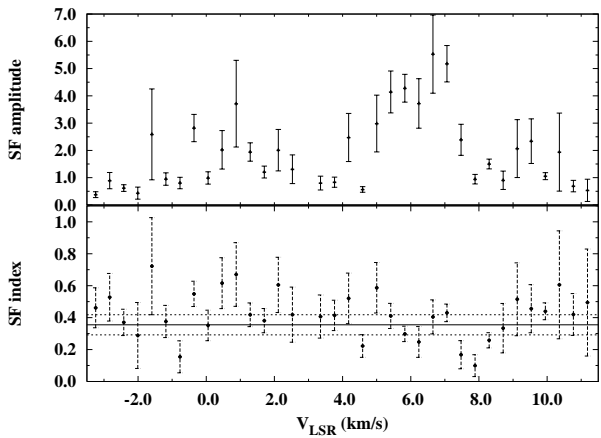}
\caption{\label{fig:ths5a} Channel to channel variation of the structure function. The best fit power law amplitude and index, with $3\sigma$ errors, are shown in the top and the bottom panel respectively. The average value of the power law index is $0.36 \pm 0.06$.}
\end{center}
\end{figure}

Apart from considering measurements from all the channels together for
fitting, each channel is considered separately to cross-check if there
is any significant channel to channel variation of the structure
function. Note that the errors on the structure function derived from
single channel reflect only the variance of the measurements, and do
not include the effect of large optical depth uncertainties. The best
fit power law amplitude and index for each channel are shown in
Figure~(\ref{fig:ths5a}). Error bars for individual channels are
$3\sigma$ error from the power law fit. The structure function
amplitude is a factor of $2-5$ higher for channels with stronger
absorption. This is consistent with the fact that the strong noise
bias in these channels are expected to scale up the
amplitude. However, we note that the average value of the power law
index $0.36 \pm 0.06$ ($3\sigma$ interval shown in dotted lines) is
consistent with no channel to channel variation of the index. This
best fit value of the power law index is also consistent (at better
than $3\sigma$ uncertainty level) with the power law index value of
$0.33 \pm 0.07$ we derived earlier from the ensemble of structure
functions from all the channels.

Typically, the power spectra from the H~{\sc i} emission and
absorption observations have a power law index $\alpha \sim 2.5-3.0$
at scales of $\sim 1000$ AU to a few pc
\citep{cro83,gre93,sta99,ddg00,beg06,dut09a,dut09b,nr10}.  For 3C~138,
the derived power law index ($\alpha \sim 2.33$) for the power
spectrum is slightly shallower. This flattening may arise from two
reasons.  Earlier it has been reported that, in the case of turbulence in
a supernova remnant, a transition from 3-d to 2-d turbulence makes the
power spectrum shallower \citep{nr09}. So, instead of homogeneous
random fluctuations, anisotropic turbulence at smaller scales can
result in a shallower spectrum. These fluctuations may be averaged out
for large scale observations with an arcsec beam. The other plausible
reason may be small scale physical processes like viscous
damping. Viscous damping for incompressible magnetized turbulence can
make the magnetic energy spectrum significantly less steep
\citep{cho02}.  Since, even with very small ionization fraction,
H~{\sc i} is coupled to the Galactic magnetic field, any possible
correlation of H~{\sc i} density with the magnetic field strength will
also flatten the opacity fluctuation power spectrum. In this regard,
it will be interesting to check if simulation of compressible
magnetohydrodynamic turbulence including damping and anisotropy at
small scales can explain these observational results quantitatively.

Our results also point out an amplitude mismatch between the observed 
structure function and that of the structure function extrapolated from large 
scale. The derived amplitude is more than an order of magnitude higher at tens 
of AU scale for 3C~138. The observed higher amplitude at small scales may be 
due to few possible reasons. First of all, for our analysis, we have used 
3C~138 ($l = 187^\circ$, $b = -11^\circ$) as a background source, whereas 
\citet{ddg00} used Cas~A ($l = 112^\circ$, $b = -2^\circ$) for deriving the 
structure function at pc scales. These two lines of sight are widely separated, 
probing H~{\sc i} in different parts of the Galaxy, where the small scale 
structures may, in principle, be intrinsically different. Alternatively, it 
may be an effect of the uncertainty in the measurement of structure function 
towards Cas A, and extrapolation over a large angular scale giving rise to 
this mismatch. In this regard, we find that extrapolating the shallower 
structure function for 3C~138 to 0.5 pc scale predicts the large scale 
structure function amplitude within $\sim 10\%$ of the observed value for 
Cas~A. In other words, the amplitude of the structure function at AU scales 
is, as expected in this second possibility, completely consistent with the 
observed rms opacity fluctuations at the pc scale. Finally, it is possible 
that there are physical mechanism(s) which modify both the amplitude and slope 
of the structure function at AU scales. For example density fluctuations due 
to intermittency may in part explain this excess of tiny scale structures. We 
note that, \citep{fal09} have reported milliparsec scale extreme velocity 
shears, which are explained as signature of intermittency. Further observations 
for more lines of sight, and measurements of structure function to bridge the 
gap between AU and pc scales will be very helpful in distinguishing between 
these different possible scenarios.

\section{Conclusions}
\label{sec:con}

We have used high resolution images of H~{\sc i} absorption towards 3C~138 
from the combined VLA, MERLIN and VLBA data to study the small scale opacity 
fluctuations. The fluctuations have a power law structure function with an 
index of $\beta \sim 0.33 \pm 0.07$ over the scales of $\sim 5-100$ AU. This 
implies a power spectrum $P_{\tau}(U) \sim U^{-2.33}$, slightly shallower than 
the earlier reported values of $\alpha \sim 2.5-3.0$ from observations over 
larger scales. The amplitude of the structure 
function at the AU scale is found to be significantly higher than the 
extrapolated amplitude from observations of Cas~A at pc scale. The observed 
rms opacity difference for 3C~138 is about $4-8$ times higher than the 
prediction from this extrapolation from Cas~A. However, extrapolating the 
observed AU scale structure function for 3C~138 consistently predicts the 
observed rms opacity fluctuations for Cas~A at the pc scale. Physical 
processes like velocity intermittency, viscous damping or anisotropy of 
turbulence may be related to this observational indication of a shallower 
slope and the presence of rich structures at smaller 
scales. Further measurements of structure function between AU and pc scales, 
and detailed numerical simulations to quantitatively verify these results 
can improve the understanding of the interstellar turbulence.

\acknowledgments

We thank the anonymous referee for many useful comments which prompted us into improving this paper substantially. We also thank John Scalo, Deputy Editor of ApJ Letters, for useful suggestions. We are grateful to Jayaram N. Chengalur, Avinash A. Deshpande, Prasun Dutta and Snezana Stanimirovi\'{c} for helpful discussions and useful comments on an earlier version of this manuscript. Part of this research was carried out at the Jet Propulsion Laboratory, California Institute of Technology, under a contract with the National Aeronautics and Space Administration.

\end{document}